\documentclass[achemso, preprint, amsfonts,amssymb]{article}
\usepackage{graphicx}
\usepackage{subfigure}

\begin{document}

\title{Validity of the ``sharp-kink approximation'' for water and other fluids}

\author{R.~Garcia\footnote{Email: garcia@wpi.edu}, K. Osborne and E. Subashi\\ 
\textit{Department of Physics, Worcester Polytechnic Institute,} \\ \textit{Worcester, MA 01609}}


%\pacs{61.30.Hn, 68.08.Bc, 68.03.Cd, 68.35.Rh, 87.15.Zg}
%\keywords{water,contact angle, wetting, electrostatics}

\maketitle

\begin{abstract}The contact angle of a liquid droplet on a solid surface is a direct measure of fundamental atomic-scale forces acting between liquid molecules and the solid surface. In this work, the validity is assessed of a simple equation, which approximately relates the contact angle of a liquid on a surface to its density, its surface tension, and the effective molecule-surface potential. This equation is derived in the sharp-kink approximation, where the density profile of the liquid is assumed to drop precipitously within one molecular diameter of the substrate. It is found that this equation satisfactorily reproduces the temperature-dependence of the contact angle for helium on alkali metal surfaces. The equation also seems be applicable to liquids such as water on solid surfaces such as gold and graphite, based on a comparison of predicted and measured contact angles near room-temperature. Nevertheless, we conclude that, to fully test the equation's applicability to fluids such as water, it remains necessary to measure the contact angle's temperature-dependence. We hypothesize that the effects of electrostatic forces can increase with temperature, potentially driving the wetting temperature much higher and closer to the critical point, or lower, closer to room temperature, than predicted using current theories.
\end{abstract}

\maketitle

\section{Introduction}
A major goal of the science of wetting and capillary phenomena is the first-principles prediction of the contact angle and/or spreading behavior of liquid droplets in contact with a surface, which may be solid or liquid. During the past two decades, there has been significant progress towards this end with the discovery and theoretical explanation of ``wetting transitions.'' These are transitions from `beading' to `spreading' that are predicted to occur quite generally as the pressure and/or temperature of a liquid on liquid or liquid on solid system is varied in the vicinity of a bulk phase transition \cite{wettingreview}. The existence of wetting transitions has been confirmed for He, H$_2$ and Ne on alkali metal surfaces at cryogenic temperatures and for Hg on sapphire and molybdenum at nearly 2000$^\circ$C \cite{prewetting}.

Wetting transitions are found to be of two types: continuous and first-order, distinguished experimentally by the behavior of the derivative of the contact angle $\theta$ near the wetting transition temperature $T_w$. Although continuous transitions, with continuous derivative everywhere, have been observed in binary mixture systems \cite{wettingreview}, to date all wetting transitions for liquid droplets on solid surfaces have been found to be first-order, were the derivative of $\theta$ has a discontinuity at $T_w$. Furthermore, in almost all experiments done to date, the temperature of the wetting transition $T_w$ has been determined not by a direct measurement of changes in the contact angle of a droplet but indirectly by a quartz microbalance or optical measurement of the prewetting transition line associated with first-order wetting transitions \cite{prewetting}. The prewetting line consists of a line of thin-thick transitions that extends out into the bulk vapor phase region from the bulk liquid-vapor coexistence line starting at $T_w$. For pressures and temperatures below the prewetting line, the adsorbed film is thin, while above the line the film is thick. For acetone on graphite, the prewetting line has been observed but the wetting transition at $T_w$ itself is unobservable because $T_w$ would lie below the triple point temperature \cite{acetone}. $^4$He on Cs \cite{ross,klier} and H$_2$ on Cs \cite{ross2} presently constitute the only two systems for which detailed measurements of the contact angle have been made all the way to $T_w$.

Under saturated vapor-pressure conditions, the equilibrium value of $\theta$ on a surface is given by the Young equation \cite{young}
\begin{equation}
\sigma_{gs}=\sigma_{ls} + \sigma_{lg} cos{ \theta} \label{eq:one}
\end{equation}
where $\sigma_{gs}$ is the gas-solid surface tension, $\sigma_{ls}$ the liquid-solid surface tension 
and $\sigma_{lg}$ the liquid-gas surface tension. In this equation both $\sigma_{gs}$ and $\sigma_{ls}$ and their temperature dependence are difficult to measure and/or determine from first principles. In fact, there exists only one instance in which it has been claimed that all four parameters in Young's equation have been independently measured \cite{israel}. Although, in this case, as expected Young's equation was satisfied within the experimental uncertainties, the ability to determine the various parameters was a fortuitous accident due to the materials involved. The impossibility of readily determining  $\sigma_{gs} - \sigma_{ls}$ for a given physical system of interest makes theoretically estimating it an important focus for the art and science of wetting theory.

The state-of-the-art theory for predicting $\sigma_{gs} - \sigma_{ls}$ presently consists of numerical simulations and mean-field density functional theory \cite{sharpkink,numerical1,numerical2}. Dietrich \cite{sharpkink}  showed that an exact expression can be obtained by finding the minimum of the grand cannonical free energy functional $\Omega[\rho(z)]$ with respect to possible density profiles $\rho (z)$ where $z$ is the distance from the substrate. The minimum of the potential, corresponding to its equilibrium value, is
\begin{equation}
\Omega_m = \sigma_{gs}- \sigma_{ls} - \sigma_{lg}\label{eq:two}
\end{equation}
so that combining eqs~\ref{eq:one} and \ref{eq:two}, one obtains
\begin{equation}
cos{\theta} = 1 + \frac{\Omega_m}{\sigma_{lg}} .\label{eq:three}
\end{equation}
Despite the simplicity of eq~\ref{eq:three}, performing the required minimization to obtain $\Omega_m$ can be quite complicated, because we must take into account the detailed microscopic fluid properties of the system.

Fortunately, in most cases, a considerable simplification is possible, which is known as the ``sharp-kink'' approximation \cite{sharpkink,wettingformula}. In this approximation, the liquid-vapor interface is assumed to have a negligible width and the liquid density is assumed to go abruptly from its bulk value to zero at a certain distance $z_{min}$ above the substrate. A series of calculations by ref~\cite{sharpkink} taking into account detailed smooth variations of $\rho (z)$ on the scale of the fluid's correlation length show that this seemingly crude ``sharp-kink'' gives surprisingly good estimates of $\Omega_m$, whenever the width over which the liquid density goes to zero at the vapor and substrate is negligible compared to the film thickness. Based on those calculations, it is found that the difficult-to-determine $\sigma_{ls}$ can be expressed very simply in terms of $\sigma_{gs}$, $\sigma_{lg}$, and the number density difference $\Delta \rho$ between liquid and vapor,
\begin{equation}
\sigma_{ls} \simeq \sigma_{gs}+\sigma_{lg} + \Delta \rho \int_{z_{min}}^{\infty} V(z)dz \,\label{eq:four}
\end{equation}
where $V=V_s-V_l$ describes the net preference of the adsorbate molecule for wetting the substrate instead of forming a droplet, due to intermolecular forces. Here $V_s$ is the potential energy of the adsorbate molecule due to the substrate and $V_l$ is the potential energy of the adsorbate molecule due to a hypothetical puddle of bulk liquid at the same location as the substrate. These terms include both attractive van der Waals forces and  repulsive forces that result from the excluded volume of the adsorbate molecule. The integral is taken over the thickness of the droplet/film starting from the position $z_{min}$ of the minimum of the potential near the surface, where $z_{min}$ is assumed to coincide with the position of the first layer of molecules on the surface.

Essentially, eq~\ref{eq:four} states that the energy cost per unit area of creating a liquid-solid interface equals the energy cost per unit area of terminating the solid in the presence of vapor plus the energy cost per unit area of terminating the liquid in the presence of vapor plus the extra potential energy per unit area gained due to the denser liquid bringing more adsorbate atoms per unit area close to the solid surface's force field. Such an approach may be expected to be valid whenever (1) changes induced in the solid by the liquid are the same as those changes in the solid induced by the vapor and (2) the lower per molecule free energy of the liquid relative to the vapor is caused solely by the denser liquid having more molecules closer to the substrate. 

By substituting eq~\ref{eq:four} into the condition $\sigma_{gs}=\sigma_{ls} + \sigma_{lg}$ for complete wetting, i.e. setting $cos\theta =1$ in eq~\ref{eq:one}, Cheng it{et al.} \cite{wettingformula} obtained a simple expression 
\begin{equation}
\frac {\sigma_{lg}} {\Delta \rho} \ge - \frac{1}{2} \int_{z_{min}}^{\infty} V(z)dz  \label{eq:five}
\end{equation}
that can be solved for $T_w$ given the temperature dependence of $\sigma_{lg}$, $\Delta \rho$ and $V$. Evidence for the validity of eq \ref{eq:four} has been found in the fact that the $T_w$ estimated from eq~\ref{eq:five} has always been within 30\% of the experimentally observed $T_w$ for all wetting transitions for which theoretical estimates are available: He, Ne, and H$_2$ on the solid alkali metal surfaces Cs and Rb at cryogenic temperatures \cite{wettingreview,wettingformula}. 

Surprisingly, however, very little has ever been said of the equation that results from simply combining eqs \ref{eq:one} and \ref{eq:four} and solving for $cos\theta$, namely
\begin{equation}
cos{\theta} = -1 -\frac{\Delta \rho}{\sigma_{lg}} \int_{z_{min}}^{\infty} V(z)dz = -1 + \frac{\Delta \rho}{\sigma_{lg}} I.\label{eq:six}
\end{equation}
For the special case of a Lennard-Jones 3-9 potential $V$ with a well-depth $D$ and van der Waals coefficient $C$, we have
\begin{equation}
V(z)=\frac {4C^3}{27D^2z^9}-\frac{C}{z^3},\label{eq:seven}
\end{equation}
which has a minimum located at $z_{min}=(2C/3D)^{1/3}$, whence the integral $I = 0.600 (CD^2)^{1/3}$. In these last two steps, we have assumed that the position of the first layer of liquid atoms near the substrate $z_{min}$ coincides with the minimum of the potential, which is reasonable in the classical limit or where the well-depth $D$ is sufficiently deep. For the special case of 3-9 potentials, numerical density functional calculations by Ancilotto it{et al.} for He on Cs yield essentially the same formula as eq \ref{eq:six} \cite{numerical2}, which demonstrates that the sharp-kink approximation is consistent with eq~\ref{eq:three} for such potentials.

The effective hard core repulsion in eq~\ref{eq:seven}, i.e. the $1/z^9$ term, and the resulting minimum in the potential are a way of accounting for the effect of the adsorbate molecule's excluded volume and the fact that an adsorbate molecule cannot approach the substrate more closely than $z_{min} \approx r_{{adsorbate}}+r_{{substrate}}$, the sum of the hard-core radii of the adsorbate and substrate molecules \cite{sharpkink}. This fact, together with $z_{min}=(2C/3D)^{1/3}$ and the fact that $r_{{substrate}}$ is generally negligible compared to $r_{{adsorbate}}$, can be used to estimate $D$ from $C$ and the average volume $V$ per molecule of the adsorbate, where, following the example of ref~\cite{colewater}, we set $\frac{4\pi}{3} r_{{adsorbate}}^3 =V$, from which we obtain
\begin{equation}
D=2C/ 3 z_{min}^3=8\pi C/3V \, . \label{eq:sevenb}
\end{equation}

Despite its apparent simplicity, eq~\ref{eq:six} has not previously been compared with it{experimentally} measured contact angles as a function of temperature $\theta(T)$. It has also never been considered as a tool for predicting contact angles from potentials or for assessing the accuracy of theoretically-computed potentials. 

In this paper, the validity of eq \ref{eq:six} is assessed using the various experimental data that are presently available, for which we also have theoretically-predicted expressions for $V(z)$. We also discuss how eq \ref{eq:six} can be used to it{predict} wetting temperatures based on empirically measured contact angles at room temperature. 

\section{Experimental Methods}

When comparing the predictions of eq \ref{eq:six} with existing contact angle data of water on graphite and gold, it was found that divergent results existed in the literature (Table 1). To investigate the source of this discrepancy, new measurements of the contact angle on these surfaces were performed. The graphite substrates used consisted of freshly-cleaved, high-quality ZYA grade graphite surfaces \cite{zya}. The gold surfaces were freshly prepared by evaporating a thin layer of Cr on a silicon wafer chip, then evaporating a micron thick gold film onto the Cr.

For the new measurements, approximately 1 mm diameter water droplets of double-distilled deionized Type I water with 8 M$\Omega$ resistitivity were used. The droplets were placed on the surface of interest under nearly saturated water vapor conditions. The use of saturated vapor conditions is very important because, in the absence of saturated vapor conditions, artificially low contact angles are obtained. The evaporating droplet exhibits a receding contact angle, which is lower than the advancing contact angle obtained under saturated vapor conditions, and depends on how quickly the droplet evaporates. 

In these measurements, negligible error was caused by the presence of air in the water-saturated atmosphere. To test this, after the droplet was placed on the surface, the surface and droplet were quickly inserted into a sealed stainless steel cell containing enough water to maintain saturated vapor conditions. The air was then removed from the cell by freezing the droplet in place, then the cell was allowed to warm back to room temperature. The contact angles of the thawed droplets in the absence of air were indistinguishable from the contact angles measured in the presence of air.

\section{Contact angle of helium and hydrogen on alkali metal surfaces}

In Fig.~1 (a), we show the contact angle data for $^4$He on substrate coated with a thick layer of Cs. The data of ref~\cite{klier} for Cs-coated W have been obtained by measuring the pressure reduction caused by capillary rise, while the data of ref~\cite{ross} for Cs-coated Au have been obtained directly from images of droplets. The solid line in the figure is the contact angle calculated using eq \ref{eq:six}. For $^4$He on Cs as well as for H$_2$ on Cs, the potential $V$ is temperature-independent. The decrease in $\theta$ with increasing temperature in Fig.~1 (a) is purely due to the divergence in the ratio ${\Delta \rho}/{\sigma_{lg}}\rightarrow \infty$ in eq \ref{eq:six} as the temperature is increased towards the critical point at $T_c=5.195$ K \cite{prewetting}.

While the theoretical line in Fig.~1 (a) based on eq~\ref{eq:six} agrees very well with the experimental data, the agreement is not perfect. The discrepancy is in part caused by the reduced $T_w$ that results from using thinner/finite thickness Cs overlayers on the Au~\cite{ross}. It was found that, as the thickness of the Cs overlayer on Au is reduced, the wetting temperature $T_w$ decreases~\cite{ross}. As shown in Fig.~1 (b), however, the discrepancy with eq~\ref{eq:six} is present even after adjusting for the reduced $T_w$ in eq~6. Empirically, the gold underlayer lowers $T_w$ from its theoretical value it{and} has the effect of multiplying the $\theta(T)$ calculated based on the lowered $T_w$ by a numerical factor that is less than unity. The W underlayer \cite{klier}, curiously, has the opposite effect, increasing the contact angle 10 \% above that predicted by eq \ref{eq:six}. Fig. 2 (b) shows the effect on the contact angle behavior of helium due to varying the Cs layer thickness on the Au substrate. Each theoretical curve is multiplied by an empirical correction factor chosen to make the curve go through the points. This correction factor allows us to quantify the magnitude of the discrepancy with eq~\ref{eq:six} for each curve. We observe that as the Cs layer is decreased from 26 to about 3 layers, where each layer is about 0.49 nm thick, the discrepancy with eq. \ref{eq:six} increases from 10\% to 25\%.

The contact angle data for H$_2$ on Cs-coated Au are shown in Fig.~2. The 30\% disagreement with theory, similar to that for $^4$He on Cs in Fig.~1, can potentially be attributed to the effect of the Au underneath the H$_2$.

\section{Contact angle of water on graphite and gold surfaces}

Given the success of eq \ref{eq:six} in describing the contact angle of helium and hydrogen on alkali metal surfaces, it is natural to wonder if it is also suitable for describing the temperature-dependence of the contact angle of more complex liquids such as water near room temperature. For hydrophobic chemically-inert surfaces like graphite, this may be the case. Dipole-dipole interactions associated with the polar nature of the water molecule are already included in the van der Waals contribution to $V$, and since hydrogen bonding does not occur between water and graphite, the cost of terminating the liquid at the vapor interface should be almost the same as near the substrate. Thus eq~\ref{eq:six} should be expected to hold to a very good approximation. 

Although, before we can accept the applicability of eq~\ref{eq:six} to water, it is crucially important to verify how the contact angle of water depends on temperature, we find much to our surprise, that very little experimental data actually exists. In Fig. 3 and Table 1, we summarize all the existing data for the contact angle of water on graphite, together together with the theoretically-predicted contact angle based on eq~\ref{eq:six}. The equilibrium contact angles reported for water are all at 25$^\circ$C. 

On the theoretical side, the potential $V(z,T)$ has recently been computed for water on a number of surfaces including graphite \cite{colewater}. Thus eq~\ref{eq:six} may be tested directly by comparing theoretically predicted and measured contact angles at 25$^\circ$C. In Fig.~3 we show $\theta(T)$ calculated using two different expressions for $V(z,T)$ for water on graphite drawn from ref~\cite{colewater}. The dashed curve is obtained using the simple 3-9 potential of eq~\ref{eq:six} with the theoretical well-depth $D=1170\pm 120$ K and van der Waals coefficient $C=12500 \pm 200$ K\AA$^3$ \cite{colewater}. The well-depth $D$ is estimated from the van der Waals coefficient $C$ and the the average volume $V$ per molecule in the bulk liquid, via eq \ref{eq:sevenb}. The solid curve is obtained using the theoretically more accurate Zhao-Johnson potential in which the 3-4-10 Steele potential is modified to include temperature-dependent dipole and quadrupole interactions \cite{zhao}. Finally, the dotted curve corresponds to choosing $I$ so that the contact angle at room temperature is 45$^\circ$. At room temperature, the simple, theoretical 3-9 potential gives a contact angle of 90.7$\pm 0.5 ^\circ$ while the more exact Zhao-Johnson potential gives an angle of 79.5$\pm 0.5 ^\circ$. The uncertainties are estimated from the stated uncertainties in $C$ and assuming a 10 \% in the theoretical value $D$ \cite{colewater}. However, it should be noted that a number of contradictory estimates of $D$ exist \cite{Grunze}. 

The symbols in Fig. 3 show the various measurements of the contact angle from Table 1 for water on graphite at room temperature. The older experimental results were done in the presence of air, while the newer measurement of 86$\pm3^{\circ}$ was done after air was removed from the cell. The new measurement in the absence of air is consistent with previously measured values of refs \cite{Fowkes} and \cite{Morcos}, and is in between the predictions of 90.8$\pm0.5 ^\circ$ based on the less rigorous 3-9 potential and 79.5$\pm0.5 ^\circ$ based on the theoretically more accurate Zhao-Johnson potential. However, the various results are inconsistent with the much lower result of ref~\cite{Schrader}. Ref~\cite{Schrader} found that surface contamination could lead to somewhat larger contact angles, and so attributed larger contact angles previously-measured to the presence of surface contaminants. However, the measurements of ref~\cite{Schrader} were apparently performed in undersaturated open air conditions and some evidence was also reported that the ion bombardment used to clean the samples damages the surfaces. Under the open air conditions of ref~\cite{Schrader}, based on our experiments and also those of ref \cite{ponter}, the evaporating droplet exhibits lower contact angles. Those results should therefore not be used. 

\begin{table*}
\begin{center}
\label{tab:one}
\caption{Experimentally measured contact angles $\theta$ of water on graphite and gold in air at 25$^{\circ}$C under various conditions. Newer measurements performed in the absence of air in this work are described in the text.} \vspace {5pt}
\begin{tabular}{l l l}
$\theta$ ($^{\circ}$) & Technique& Reference\\
\hline
$85.6 \pm 0.3 $& Optical obs., ZYH graphite& (1940)\cite{Fowkes}\\
$83.9 \pm 0.1$& Capillary rise, ZYH graphite& (1972)\cite{Morcos}\\
$46 \pm 3 $& Optical obs., freshly cleaved ZYA graphite&(1980)\cite{Schrader}\\
$35.9 \pm 0.8 $& Optical obs., Electron Bombardment&(1980\cite{Schrader}\\
$86 \pm 3 $& Optical obs., freshly cleaved ZYA graphite&(2007) This work\\
\hline 
$ 52-78$& Optical obs., Gold & (1948)\cite{plaksin}\\
$ 62.6\pm 3.4 $& Optical obs., Gold & (1965)\cite{Erb}\\
$ 55-65$& Optical obs., Gold & (1966)\cite{White}\\
$ 56.6 \pm1.5$& Optical obs., Freshly evaporated Gold &(2007) This work\\
\hline
\end{tabular}
\end{center}
\end{table*}

From Fig.~3, we see that the decrease in the contact angle for water on graphite is expected to become significant only above $\sim$ 120 $^{\circ}$C, where experiments are more difficult because of high vapor pressures and increased reactivity of water. Nevertheless, accurate predictions are also available for water on Au, where using the theoretically-predicted $C=19100 \pm 400$ K \AA$^3$ from ref~\cite{colewater} we calculate $D=1790\pm 180$ K, obtaining a similar curve to that of Fig.~2, which in this case gives $\theta=59.2\pm0.9 ^{\circ}$ at 25$^{\circ}$C and $T_w=143\pm3 ^\circ$C. The experimentally-observed contact angles (Table 1) on average agree with the predicted contact angle at room temperature. For our measurements, we used a freshly evaporated gold surface. As with water on graphite, however, experiments have not yet been done to verify the predicted temperature-dependence. 

\section{Other liquids and surfaces}

Even if we lack detailed knowledge of the surface-adsorbate attractive potential, eq~\ref{eq:five} can still be used to roughly predict which systems are likely to exhibit wetting transitions near room temperature, based only on the contact angle $\theta_0$ that the liquid exhibits at some temperature $T_0$. First we estimate the integral $I$, which itself is not very temperature sensitive by substituting $\theta_0$, $\Delta \rho (T)$ and $\sigma_{lv}(T)$ into eq~\ref{eq:six}; then we find $T_w$ by subtituting $\theta= 0$ and $I$. Based on the room-temperature contact angles on teflon of $43.70\pm 0.06^{\circ}$ for decane and $47.96\pm0.10 ^{\circ}$ for dodecane \cite{contactangles}, we are able to predict  $T_w = 71.0\pm0.5^{\circ}$C for decane and $T_w = 86.5\pm0.5 ^{\circ}$C for dodecane. These temperatures are much lower than for water on graphite; thus it may be easier to confirm the validity of eq~\ref{eq:six} near room temperature using these nonpolar liquids. 

\section{Electrostatic effects}

Based on eq~\ref{eq:six}, for water to exhibit a wetting transition below 100 $^{\circ}$C on a given surface, it would have to exhibit a contact angle of $45^{\circ}$ or less at 25 $^{\circ}$C. However, it is important to realize that, within the scheme of eq~\ref{eq:six}, we are able to accurately predict $T_w$ solely from the room temperature contact angle it{only if} the substrate-induced potential energy gain per unit area is proportional to the adsorbate number density $\rho$. But two important types of forces can be expected to violate this assumption: pinning forces and electrostatic forces. 

Pinning forces are generally acknowledged for their nonlinearity and are difficult to model theoretically \cite{wettingreview}. Electrostatic forces in general also are not negligible. This can be shown using a simple expression for the electrostatic pressure used previously by Langmuir to explain the anomalously thick water films on glass \cite{lang}
\begin{equation}
P_e=\frac{\epsilon \epsilon_o}{2z^2} \left( \frac{\pi k_B T}{e} \right)^2 \sim \frac{1.5 \times 10^{-14}{N}}{z^2}\, ,\label{eq:eight} 
\end{equation}
where for water $\epsilon \sim 80$, and $T=300$ K, and $e$ is the elementary electric charge. This equation applies in the limit of large film thickness and large surface charges. Comparing $P_e$ to the pressure caused by the van der Waals attraction of water to graphite
\begin{equation}
P_v  \sim \frac{\rho_l \, C}{z^3} \sim 1.4 \, \frac{k_B T}{z^3}.\label{eq:eight} 
\end{equation}
one obtains that $P_e$ exceeds $P_v$ for $z > 1$ nm. This calculation shows that the surface tension contribution from electrostatic forces can easily dominate over the contribution from van der Waals forces under certain circumstances. Nevertheless, this represents a limiting worst-case, because graphite and water are known not to react under ordinary circumstances, so severe charging of the interface is not expected. 

In the case of a charged droplet under constant charge conditions, potentially applicable to many experiments, it has been predicted \cite{kang} that at least two extra electrostatic terms must be added, so that eq~\ref{eq:six} would read
\begin{eqnarray}\label{eq:nine}
cos{\theta} = -1 -\frac{\Delta \rho}{\sigma_{lg}} \int_{z_{min}}^{\infty} V(z)dz \\ \nonumber
+ \frac {k_B T} {\sigma_{lg}} A(\phi_{\infty}, n, T) - \frac {\Sigma} {\sigma_{lg}}  \left(\phi_0-\phi_{\infty}\right), 
\end{eqnarray}
where $A$ is a function of the electrostatic potential $\phi_{\infty}$ in the bulk far away from the contact line, $n$ the number density  of the it{ion} species in the droplet, $\Sigma$ the charge density at the substrate interface, and $\phi_0$ the potential at the contact line. 

Thus depending on the precise circumstances, which determine the magnitude and sign of the last two terms in eq~\ref{eq:nine}, electrostatic forces can favor either wetting or non-wetting. Further, these forces are independent of $\Delta \rho$ and may increase rather than decrease in effect as the temperature is increased.  To validate eq~\ref{eq:six} it is, therefore, important to select an unreactive surface which does not spontaneously become charged when coming in contact with the liquid and one that is at least sufficiently conducting that accidental static charging can be dissipated. Graphite and gold in most respects seem to be ideal candidates, although some electrostatic effects may still be present \cite{yaminski}. Finally, we must also note that for layered substrates the observed contact angles appear to be smaller than predicted from eq~\ref{eq:six}; this would lead us to underestimate $T_w$ based on a given $\theta_0$.

One of the key predictions of the Cahn theory of wetting is that $\theta(T) \rightarrow 0$ due to $\Delta \rho/\sigma_{lg} \rightarrow \infty$ as the critical point is approached. Equation~\ref{eq:nine}, however, is significant because the last two terms are not proportional to $\rho$. We only recover the Cahn prediction if the last two terms tend to zero as we approach the critical point. This assumption is reasonable but not immediately obvious. Furthermore, even if the last two terms vanish near the critical point, the predicted $\theta(T)$ can in principle significantly differ from the $\theta(T)$ eq~\ref{eq:six}. In the presence of charge, the estimated wetting temperature can be much higher or lower than predictions made solely on the basis of eq~\ref{eq:six} \cite{colewater}.

\section{Conclusions}

In addition to eq~\ref{eq:three}, various alternatives to eq~\ref{eq:six} have previously been proposed \cite{alternatives}. In terms of the disjoining pressure $\Pi$, it is possible to write $cos{\theta} = 1 + \int_{z_{min}} \Pi(z)dz $. Using one very common empirical approximation, $cos{\theta} = -1 + 2\left( \gamma^d_{l} \gamma^d_{s} \right)^{1/2}/\gamma_{l} $ where the $\gamma$'s are empirical effective free-energy parameters associated with the solid and liquid phases. Although these alternatives are intended to encompass the same physics, they lack the simplicity of interpretation of eq~\ref{eq:six}. Its straight-forward meaning may be clearly seen by considering three cases: (i) in the absence of an attraction to the substrate ($V(z)=0$), we have complete non-wetting $\theta= \pi$; (ii) as the attractive potential becomes stronger (more negative), $\theta$ decreases monotonically to zero; (iii) for a sufficiently strong attractive potential, we have complete wetting $\theta= 0$. Furthermore, we have shown that eq~\ref{eq:six} provides a transparent, quantitative description of how the contact angle of a liquid decreases as we increase the temperature towards its critical point along the liquid-vapor coexistence line.

Eq~\ref{eq:six} accurately predicts the contact angle of $^4$He on Cs as a function of temperature for sufficiently-thick Cs substrates. For Au substrates coated with a very thin Cs layer, however, eq~\ref{eq:six} overestimates the contact angle by as much as 30\%, the discrepancy, in both $T_w$ and the magnitude of the predicted contact angle, being greater for thinner Cs films. Further work is necessary to identify a theoretical basis for this effect. 

Preliminary analysis of the contact angle of water on graphite and gold at room temperature suggests that eq~\ref{eq:six} accurately predicts the contact angle of water on these substrates and therefore may be suitable for predicting the wetting temperature of water and other liquids on a given substrate based on only their contact angles at room temperature, whenever pinning and electrostatic forces are negligible. 

Assuming the validity of eq~\ref{eq:six} it is predicted that for water to exhibit a wetting transition below 100 $^{\circ}$C on given surface, it would have to have had a contact angle on the order of $45^{\circ}$ at 25 $^{\circ}$C. $T_w$ for decane and dodecane on teflon are expected to be 71 and 87 $^{\circ}$C, respectively, making them ideal candidates for testing the theory for nonpolar liquids near room temperature. 

Nevertheless, because charging effects add new terms to eq~\ref{eq:six} that are difficult to quantify a priori, experimental measurement of $\theta(T)$ is necessary, before a firm conclusion can be made regarding the applicability of eq~\ref{eq:six} without those additional terms. 

When making our contact angle measurements for water on graphite, no special efforts were made to clean the graphite other than to use a freshly cleaved sample.  Similarly for gold surfaces we merely tested contact angle on freshly evaporated gold films.  Future improvements should include annealing the graphite under an argon atmosphere and examining the effect of water purity \cite{hess}. 

\section{Acknowledgment}
We acknowledge useful conversations with M. W. Cole, S. Dietrich, G. Iannacchione and M. Fukuto. This research was supported by PRF grant 45840-G5.
%%%%%%%%%%%%%%%%%%%%%%%%%%%%%%%%%%%

%%%%%%%%%%%%%%%%%%%%%%%%%%%%%%%%%%%
\begin{figure}[ht]
\vskip 0.1 cm \includegraphics[angle=0,width=90mm]{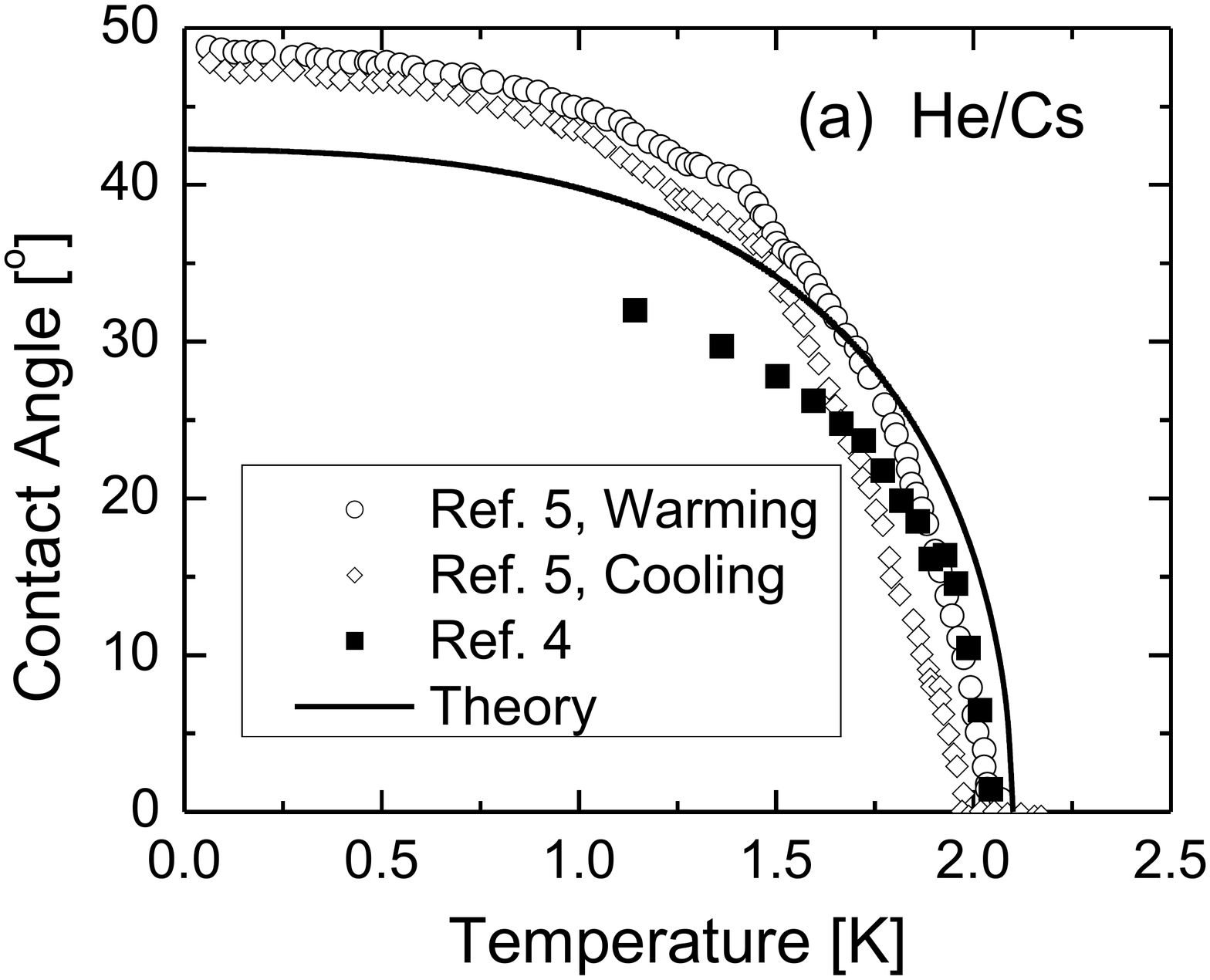} 
\vskip 0.1 cm \noindent
\vskip 0.1 cm \includegraphics[angle=0,width=90mm]{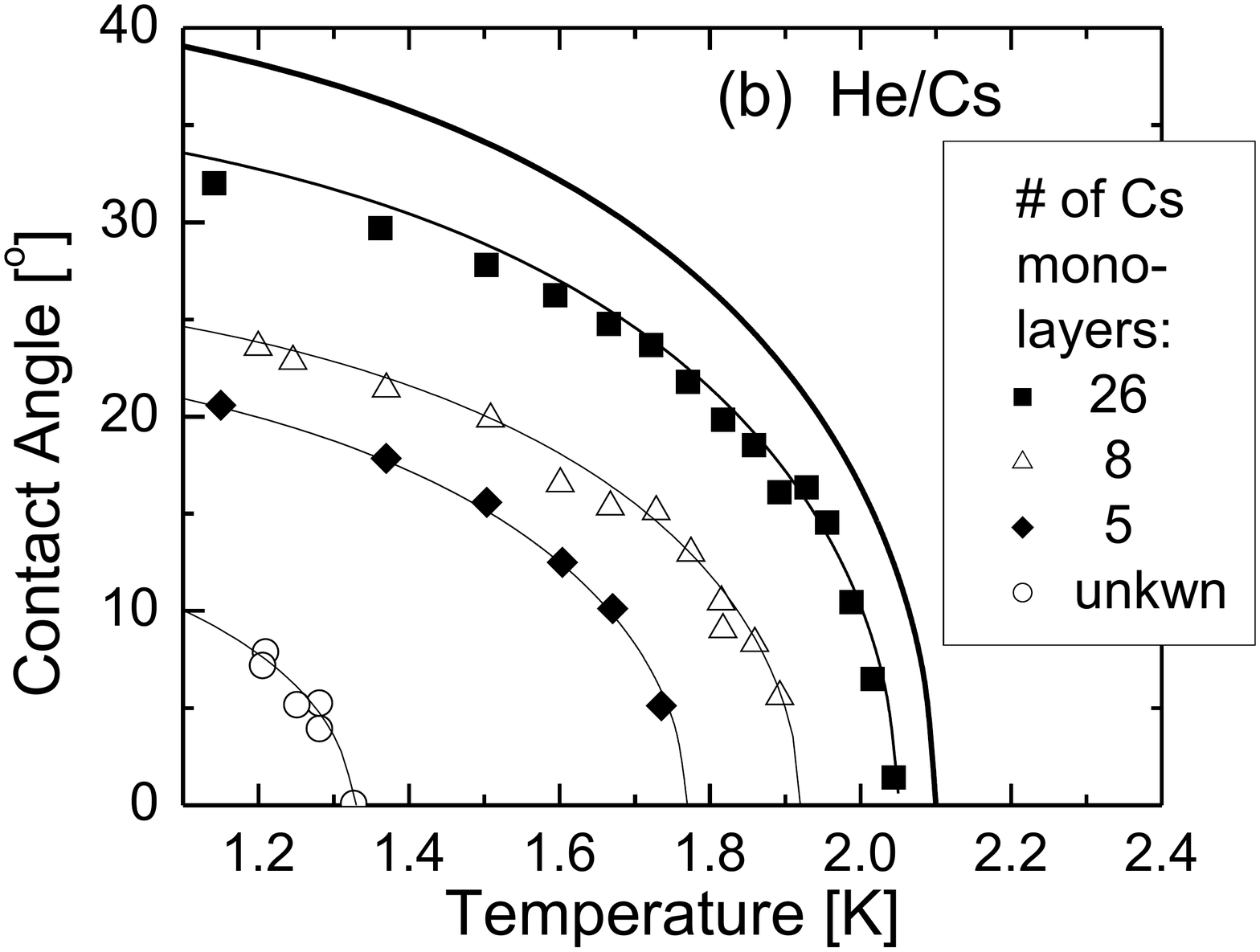}
\vskip 0.1 cm \noindent
\caption{(a) Measured $\theta (T)$ for $^4$He on Cs \cite{ross,klier} \textit{vs.} $\theta (T)$ predicted  from eq~\ref{eq:six} assuming a value of $I=20.4$~K\AA~ from ref~\cite{heliumparameters}.
(b) Effect on $\theta (T)$ of varying the thickness of the Cs coating on the Au from ref~\cite{ross}. The number of Cs monoloayers deposited on the Au was 26, 8, 5 and thin-but-uncertain, respectively. Thinner films showed a lower $T_w$ and deviated more significantly from the the theory. The thick solid line is the predicted $\theta (T)$ from Fig. 1(a); the thin lines are the predicted $\theta (T)$ for the surface, based on the experimentally measured $T_w =$ 2.06, 1.92, 1.77 and 1.32 K from ref~\cite{ross}, multiplied by factors of 0.90, 0.75, 0.75, and 0.75, respectively, so that the curves go through the points. }
\label{fig1} 
\end{figure}

\begin{figure}[ht]
\vskip 0.1 cm \includegraphics[angle=0,width=90mm]{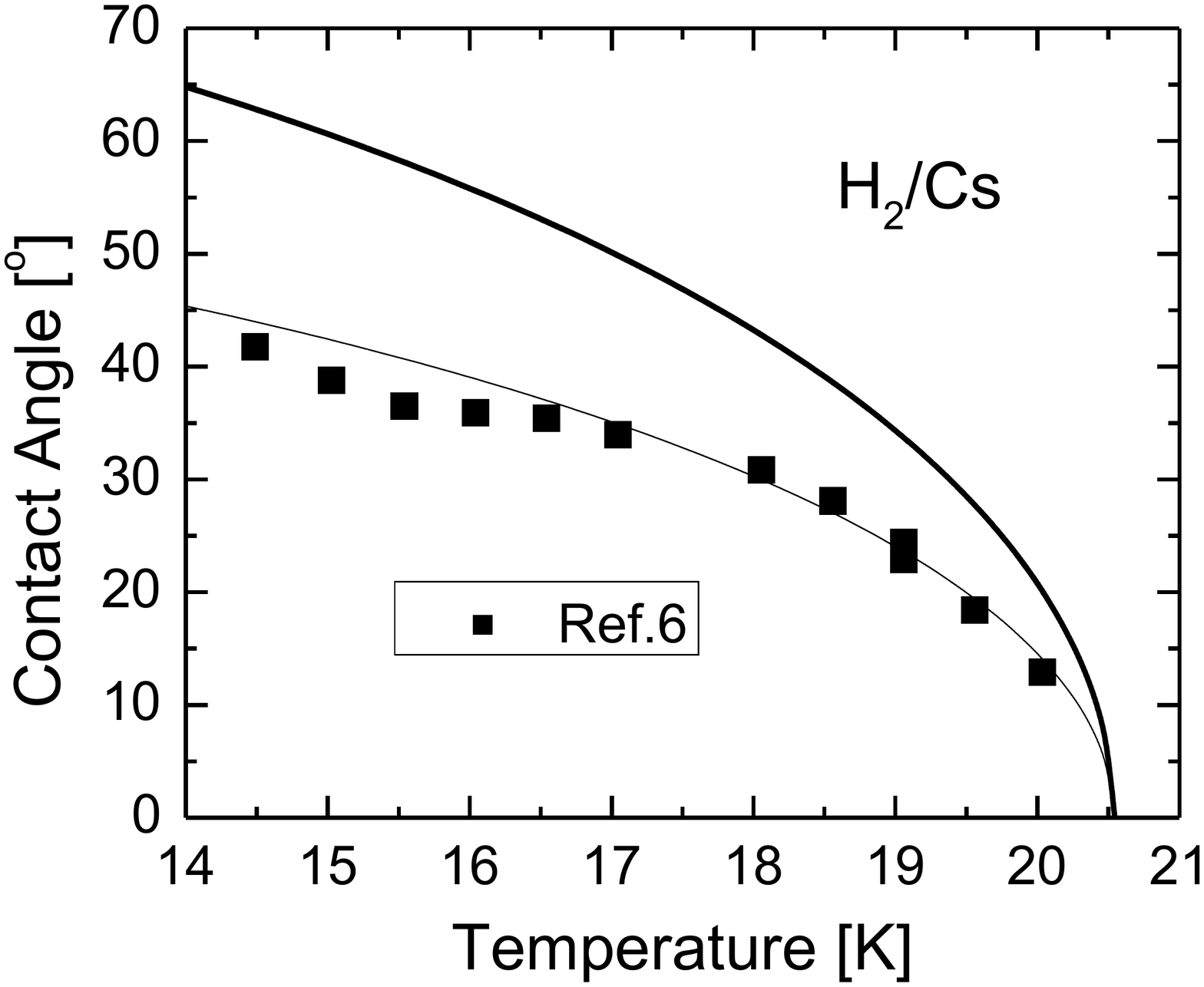}
\vskip 0.1 cm \noindent
\caption{Measured $\theta (T)$ for H$_2$ on Cs \cite{ross2} \textit{vs.} (thick line) $\theta (T)$ predicted from eq~\ref{eq:six} using $I=133.9$~K\AA~ from ref~\cite{ross2}. The thin line going through the points is the same predicted $\theta (T)$ multiplied by a correction factor of 0.70.}
\label{fig2} 
\end{figure}

\begin{figure}[ht]
\vskip 0.1 cm \includegraphics[angle=0,width=90mm]{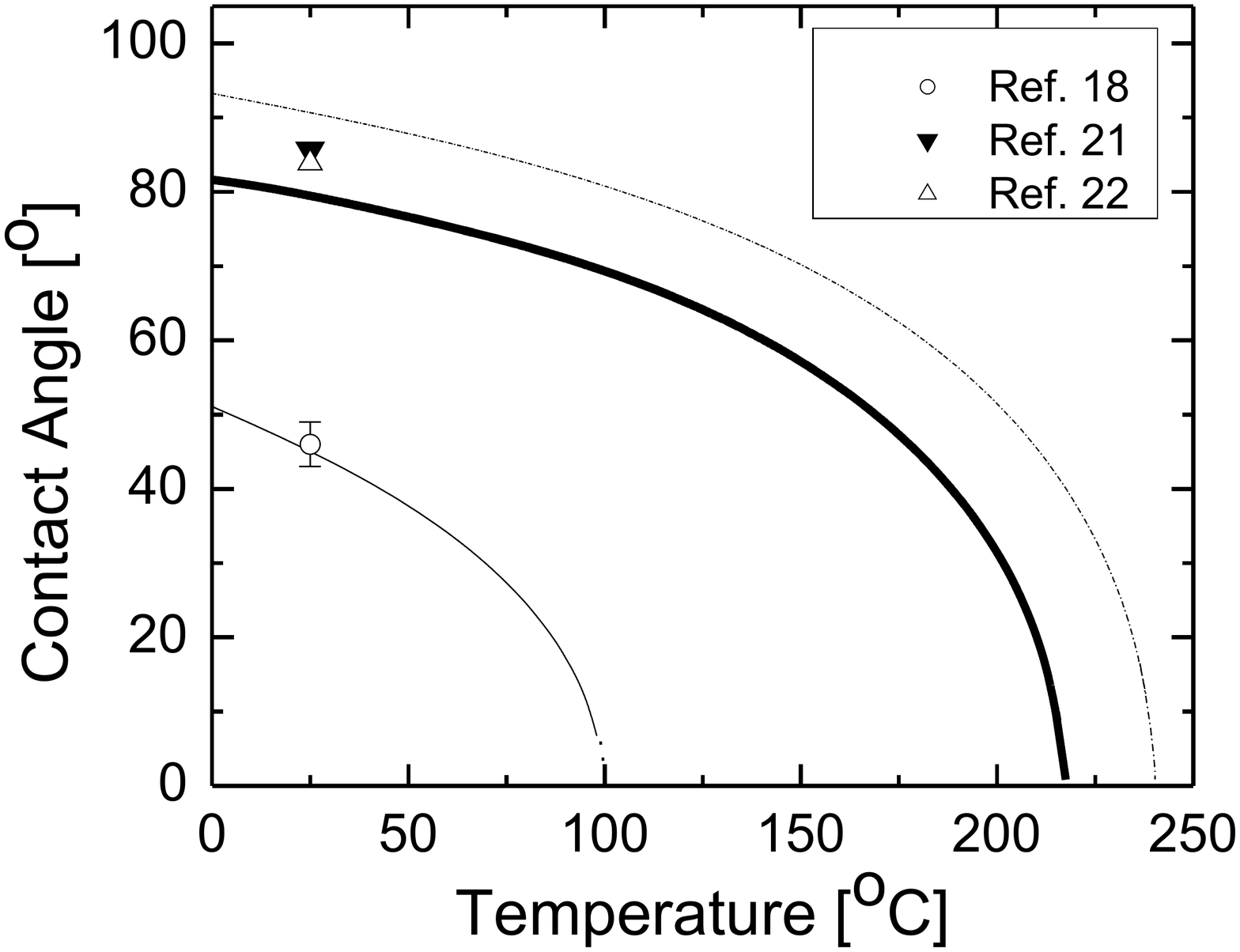} 
\vskip 0.1 cm \noindent
\caption{Predicted $\theta (T)$ for water on graphite. The thin, upper curve is obtained using the simple 3-9 potential of eq~\ref{eq:seven} and theoretically-predicted $D=1170\pm 120$ K and $C=12500 \pm 1300$ K \AA$^3$ \cite{colewater}. The thick curve in the middle is obtained using the theoretically more accurate Zhao-Johnson potential \cite{zhao}. The thin, lower curve corresponds to a hypothetical substrate for which $\theta = 45^\circ$ at room temperature. Symbols show experimentally measured contact angles from Table 1. For clarity we do not show the new data point obtained from this work because it overlaps the data from Ref. 22 and 21.}
\label{fig3} 
\end{figure}

\begin{figure}[ht]
\vskip 0.1 cm \includegraphics[angle=-90,width=80mm]{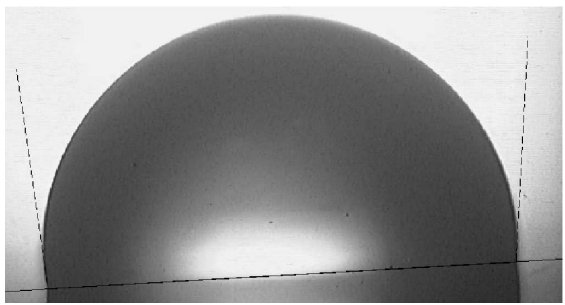} 
\vskip 0.1 cm \noindent
\caption{Typical photograph of a 2 $\mu$L water droplet on ZYA graphite at 25$^{\circ}$C under saturated conditions. The width of the droplet is 2.0 mm.}
\label{fig4} 
\end{figure}

\end{document}